\documentclass[aip,jcp,reprint,showpacs,amsmath,amssymb,longtable]{revtex4-2}
\usepackage{graphicx}
\usepackage{latexsym}
\usepackage{amsmath}
\usepackage[english]{babel}
\usepackage{amssymb}
\usepackage{amsfonts}
\usepackage{physics}
\usepackage{makecell}
\usepackage{longtable}
\usepackage{multirow}
\usepackage{subfigure}
\usepackage{xcolor}
\usepackage[utf8]{inputenc}   
\allowdisplaybreaks

\begin{document}

\title{Double excitations in molecules}

\author{Namana Venkatareddy}
\affiliation{Centre for Condensed Matter Theory, Department of Physics, Indian Institute of Science, Bangalore 560012, India}
\author{Victor Ghosh}
\affiliation{Centre for Condensed Matter Theory, Department of Physics, Indian Institute of Science, Bangalore 560012, India}
\author{H. R. Krishnamurthy}
\affiliation{Centre for Condensed Matter Theory, Department of Physics, Indian Institute of Science, Bangalore 560012, India}
\affiliation{International Centre for Theoretical Sciences, Tata Institute of Fundamental Research, Bengaluru 560089, India}

\author{Manish Jain}
\affiliation{Centre for Condensed Matter Theory, Department of Physics, Indian Institute of Science, Bangalore 560012, India}

\date{\today}

\begin{abstract}
Double excitations in organic molecules have garnered significant interest as a result of their importance in singlet fission and photophysics. These excitations play a crucial role in understanding the photoexcitation processes in polyenes. To describe photoexcited states with both single and double excitation character, we use a first-principles many-body theory \cite{PhysRevB.100.201403} that combines the GW / Bethe-Salpeter equation and the configuration interaction (CI) methods. Specifically, we develop and employ two CI-based methods: screened configuration interaction singles and doubles (scrCISD) and screened configuration interaction singles with perturbative doubles (scrCIS(D)), applied to an effective many-body Hamiltonian \cite{PhysRevB.100.201403} that incorporates screening. We apply these methods to Thiel's set of molecules \cite{10.1063/1.2889385}, which exhibit excited states predominantly characterized by single excitations with a partial double excitation character. Our results indicate that the scrCISD method systematically underestimates the excitation energies compared to the best theoretical estimates, while the scrCIS(D) method shows good agreement with these estimates. Furthermore, we used the scrCISD method to calculate the binding energies of the dominantly doubly excited correlated triplet pair states\cite{doi:10.1021/acs.jpca.5b09725}, $\mathrm{TT^1}$, in pentacene dimers, finding that the $\mathrm{TT^1}$ binding energies agree well with empirical calculations. 
\end{abstract}

\maketitle

\section{Introduction}
Double excitations in molecules play an important role in several processes, such as singlet fission and photophysics. Singlet fission\cite{doi:10.1021/cr1002613}, which has applications in improving solar cell efficiency, involves an intermediate state that is predominantly doubly excited.  In certain polyenes \cite{HUDSON1984241,starcke2006much} and carotenoids \cite{frank1996carotenoids}, low-lying electronic states exhibit a partial double excitation character, which plays a crucial role in the photophysics of these molecules. Thus, studying and understanding the role of double excitations in the excited states of molecules has become increasingly essential. 

Accurately describing double excitations requires computationally expensive high-level wave function methods\cite{doi:10.1021/acs.jctc.8b01205}. These include coupled cluster approaches such as coupled cluster singles and doubles with perturbative triples method\cite{10.1063/1.470315} (CC3), coupled cluster singles, doubles and triples method\cite{10.1063/1.452353} (CCSDT) and coupled cluster singles, doubles, triples and quadruples method\cite{Kucharski1991} (CCSDTQ). Additionally, multi-configurational methods like complete active-space self-consistent field\cite{doi:https://doi.org/10.1002/9780470142943.ch7} (CASSCF), complete active-space second-order perturbation theory\cite{article} (CASPT2) and single-state N-electron valence state perturbation theory\cite{ANGELI2001297} (NEVPT2) are also used. Within the framework of their most commonly used approximation, \textit{i.e.} frequency-independent kernels, popular single excitation methods such as time-dependent density functional theory (TDDFT)\cite{Runge1984997} and the GW-Bethe-Salpeter equation (GW-BSE)\cite{PhysRevB.62.4927} fail to adequately describe double excitations. TDDFT is often solved using the adiabatic approximation, where a frequency-independent exchange-correlation kernel is employed. Furthermore, GW-BSE usually employs a static screened Coulomb interaction instead of a dynamically screened Coulomb interaction. The use of frequency-independent kernels results in the absence of double (and higher) excited states in the calculated spectrum\cite{10.1063/5.0028040}.

In the context of TDDFT, several approaches have been developed to go beyond the standard adiabatic approximation. For example, Casida\cite{10.1063/1.1836757} proposed a nonadiabatic correction to the exchange-correlation kernel using the superoperator formalism. Maitra\cite{10.1063/1.1651060} \textit{et al.} introduced the dressed TDDFT approach. They derived an exact frequency-dependent exchange-correlation kernel for cases where a double excitation is mixed with a single excitation and well-separated from the rest of the spectrum. For a more general case, Romaniello\cite{10.1063/1.3065669} \textit{et al.} used the Bethe-Salpeter equation with a dynamically screened Coulomb interaction $W(\omega)$ to construct a frequency-dependent exchange-correlation kernel for TDDFT.

Similarly, in the case of GW-BSE, several approaches have been used to include the dynamically screened Coulomb interaction. Ma\cite{PhysRevB.80.241405} \textit{et al.} perturbatively included the effects of dynamical screening within the plasmon pole approximation.    Bintrim\cite{10.1063/5.0074434} \textit{et al.} demonstrated that the full-frequency GW-BSE calculation can be reformulated as a frequency-independent eigenvalue problem in the expanded subspace of single and double excitations. This approach allows one to solve a larger linear eigenvalue problem rather than a nonlinear eigenvalue problem on a smaller subspace, as is done with a frequency-dependent kernel. Analogously, our formalism also uses frequency-independent Hamiltonian and explicitly includes double excitations in the basis. 

We use a first-principles many-body theory \cite{PhysRevB.100.201403} that combines the GW-BSE approach with the configuration interaction (CI) method to describe doubly excited states in molecules. Our method involves an effective many-body Hamiltonian\cite{PhysRevB.100.201403} that incorporates screening. Various CI methods have been applied to this Hamiltonian to describe excitations in solids, such as trions\cite{PhysRevB.100.201403} and biexcitons\cite{Torche2021}. Specifically, to address double excitations in molecules, we employ two CI-based methods: screened configuration interaction singles and doubles (scrCISD) and screened configuration interaction singles with perturbative doubles (scrCIS(D)). These methods are applied to a benchmark set of molecules known as Thiel's set\cite{10.1063/1.2889385}, which contains excited states that are predominantly singly excited with partial contributions from double excitations. The performance of our methods on Thiel's set is evaluated by comparing their outcomes against the best theoretical estimates. Additionally, the scrCISD method is used to calculate the binding energies of dominantly doubly excited correlated triplet-pair states $\mathrm{(TT)}^1$.

The paper is organized as follows: Section II provides a detailed description of our theory. Section III outlines the computational details of the calculations on Thiel's set and dimer molecules. Section IV presents the results and their implications. Finally, Section V provides the concluding discussion.

\section{Theory}
Our formalism for describing double excitations is based on the first-principles many-body theory proposed by A. Torche and G. Bester\cite{PhysRevB.100.201403}. They use an effective many-body Hamiltonian that incorporates screening. The effective Hamiltonian $H$ is given by 
\begin{align}
 H & = H_{0} + H_{ee} + H_{hh} + H_{eh} + H_{x}  \nonumber \\
 H_0& = \sum_{i} \epsilon_{i}^{qp}a_{i}^{\dagger}a_{i} - \sum_{\alpha} \epsilon_{\alpha}^{qp} b_{\alpha}^{\dagger}b_{\alpha} \nonumber \\
 H_{ee} & = \frac{1}{2}\sum_{i,j,k,l} \bra{ij}W\ket{lk}a_{i}^{\dagger}a_{j}^{\dagger}a_{k}a_{l} \nonumber \\
 H_{hh} & = \frac{1}{2}\sum_{\alpha,\beta,\gamma,\delta} \bra{\alpha \beta} W \ket{\gamma \delta}b_{\delta}^{\dagger}b_{\gamma}^{\dagger}b_{\alpha}b_{\beta} \nonumber \\
H_{eh} & = \sum_{i,j,\alpha,\beta}[\bra{i \beta}{v}\ket{ \alpha j}- \bra{i \beta}W\ket{j \alpha }] C_{i \alpha}^{\dagger}C_{j \beta} \nonumber \\
H_{x} & =H_{xx} + H_{xe} + H_{xh} \nonumber \\
H_{xx} & = \frac{1}{2} \sum_{i,j,\alpha,\beta} \bra{ij}{W}\ket{\alpha \beta}_{as}[ C_{i \alpha}^{\dagger}C_{j \beta}^{\dagger}+  C_{i \alpha}C_{j \beta}] \nonumber \\
H_{xe} & = \frac{1}{2} \sum_{i,j,k,\alpha} \bra{ij}{W}\ket{k \alpha}_{as}[ C_{j \alpha}^{\dagger}a_{i}^{\dagger}a_{k}+  a_{k}^{\dagger}a_{i}C_{j \alpha}] \nonumber \\
H_{xh} & = \frac{1}{2} \sum_{\alpha,\beta,\gamma,i} \bra{\alpha\beta}{W}\ket{\gamma i}_{as}[ C_{i \alpha}^{\dagger}b_{\beta}^{\dagger}b_{\gamma}+  b_{\gamma}^{\dagger}b_{\beta}C_{i \alpha}]
\end{align}
In the above expressions, we use $i,j,k$ to represent electron states (one particle states that are higher in energy than the Fermi energy) and $\alpha,\beta,\gamma$ to represent hole states (one particle states that are lower in energy than the Fermi energy).  $a_i, a_i^{\dagger}(b_{\alpha}, b^{\dagger}_{\alpha})$ are electron (hole) creation and annihilation operators, respectively, and $C_{i \alpha}^{\dagger} = a_{i}^{\dagger}b_{\alpha}^{\dagger}$. $\epsilon^{qp}$ denotes quasiparticle energies from
a GW calculation. The screened Coulomb interaction, $W$, in terms of inverse dielectric function $\epsilon^{-1}$ is given by.
\begin{align}
W(r_1,r_2) &= \int \epsilon^{-1}(r_1,r_3) v(r_3,r_2) dr_3     
\end{align}
The matrix elements corresponding to screened ($W$) and bare ($v$) Coulomb interactions are calculated from single particle states $\phi_{i}$ calculated from a mean-field calculation. 
\begin{align}
\bra{ij}W\ket{kl} &= \int dr_{1}dr_{2}\phi^{*}_{i}(r_1) \phi^{*}_{j}(r_2)W(r_1,r_2)\phi_{k}(r_1) \phi_{l}(r_2) \nonumber\\
\bra{ij}W\ket{kl}_{as} &= \bra{ij}W\ket{kl} -\bra{ij}W\ket{lk} \nonumber \\  
\end{align}
$H_{0}$ is the non-interacting part of the Hamiltonian (containing energies of free electrons and holes). $H_{ee}$ and $H_{hh}$ describe the quasi-electron and quasi-hole interactions (screened), respectively. The term $H_{eh}$ describes electron-hole interactions that are equivalent to the hybridization of $e-h$ excitations that lead to the formation of excitonic bound states. It contains attractive direct interaction terms and repulsive exchange interaction terms. As in the GW-BSE Hamiltonian, the electron-hole direct interaction is screened, whereas the exchange interaction is left unscreened. The $H_{xe}$, $H_{xh}$, and $H_{xx}$ terms lead to exciton-electron interaction, exciton-hole interaction, and the creation and destruction of pairs of excitons, respectively.

The inclusion of screening corresponds to the renormalization of the Coulombic interaction between electrons and holes in low-lying excitations, such as excitons, trions, and so on. This renormalization occurs due to their interaction with high-energy charge-density fluctuations, namely plasmons\cite{PhysRev.139.A796}.     

The configuration interaction (CI) method was developed to describe various excitations of the effective many-body Hamiltonian. In the CI method, excitation energies of a specific type of excitation (single or double) are calculated by projecting the Hamiltonian onto the corresponding restricted subspace and then diagonalizing it. For example, for dominantly singly excited excitations, the projection subspace consists of a set of all singly excited Slater determinants. The projection subspace $P^{x}$ in case of single excitation is given by 
\begin{equation}
   P^{x} = \text{Span}\{ \ket{i\alpha}, \ket{i\alpha} = a_{i}^{\dagger}b_{\alpha}^{\dagger}\ket{\Omega}\} 
\end{equation}
In the above expression, $\ket{\Omega}$ is the non-interacting ground state Slater determinant. The matrix representation for the projected Hamiltonian for single excitation $H^{x}$ is given by 
\begin{align*}
    H_{j\beta,i\alpha}^{x} &= \bra{i\alpha}H\ket{j\beta} \nonumber
\end{align*}
\begin{align}
   \bra{i\alpha}H\ket{j\beta} &= \left(\epsilon_i^{qp} - \epsilon_\alpha^{qp}\right) \delta_{ij}  \delta_{\alpha \beta} + \mel{i \beta}{v}{\alpha j} - \mel{i \beta}{W}{j \alpha}
\end{align}
The eigenvalue problem for the Hamiltonian $H^{x}$ is the same as the standard GW-BSE equation within the Tamm-Dancoff approximation (TDA). Diagonalizing the projected Hamiltonian $H^{x}$ thus yields excited states corresponding to single excitations.

For dominantly doubly excited states, the projection subspace corresponds to a set of all doubly excited Slater determinants. The projection subspace $P^{xx}$ is given by
\begin{equation}
P^{xx} = \text{Span}\{ \ket{ij\alpha \beta} ,\ket{ij\alpha \beta} = a_{i}^{\dagger}a_{j}^{\dagger}b_{\alpha}^{\dagger}b_{\beta}^{\dagger} \ket{\Omega}, i>j,\alpha > \beta \} 
\end{equation}
The projected Hamiltonian matrix for double excitations $H^{xx}$ is given by
\begin{align}
    H_{mn\gamma\eta,ij\alpha\beta}^{xx} = &\bra{ij\alpha\beta}H\ket{mn\gamma\eta} \nonumber \\
    \bra{ij\alpha\beta}H_0\ket{mn\gamma\eta}  = & (\epsilon^{qp}_i + \epsilon^{qp}_j - ( \epsilon^{qp}_{\alpha} + \epsilon^{qp}_{\beta}) ) \delta_{im}\delta_{jn}\delta_{\alpha \gamma}\delta_{\beta \eta} \nonumber \\
 \bra{ij\alpha\beta}H_{ee}\ket{mn\gamma\eta}  = &[\bra{ij}W\ket{mn}-\bra{ij}W\ket{nm}]\delta_{\alpha \gamma} \delta_{\eta \beta} \nonumber \nonumber  \\ 
\bra{ij\alpha\beta}H_{hh}\ket{mn\gamma\eta}  = & [\bra {\gamma \eta} W \ket{ \alpha\beta}-\bra{ \gamma \eta} W \ket{\beta\alpha }]\delta_{m i} \delta_{n j} \nonumber \\ 
 \bra{ij\alpha\beta}H_{eh}\ket{mn\gamma\eta} = &  -\left[ \bra{i \eta}v\ket{\alpha m} - \bra{i \eta}W\ket{m \alpha} \right] \delta_{jn} \delta_{\beta \gamma} \nonumber \\
& + \left[ \bra{i \eta}v\ket{\alpha n} - \bra{i \eta}W\ket{n \alpha} \right] \delta_{jm} \delta_{\beta \gamma} \nonumber \\
& + \left[ \bra{j \eta}v\ket{\alpha m} - \bra{j \eta}W\ket{m \alpha} \right] \delta_{in} \delta_{\beta \gamma} \nonumber \\
& - \left[ \bra{j \eta}v\ket{\alpha n} - \bra{j \eta}W\ket{n \alpha} \right] \delta_{im} \delta_{\beta \gamma} \nonumber \\
& - \left[ \bra{i \gamma}v\ket{\beta m} - \bra{i \gamma}W\ket{m \beta} \right] \delta_{jn} \delta_{\alpha \eta} \nonumber \\
& + \left[ \bra{i \gamma}v\ket{\beta n} - \bra{i \gamma}W\ket{n \beta} \right] \delta_{jm} \delta_{\alpha \eta} \nonumber \\
& + \left[ \bra{j \gamma}v\ket{\beta m} - \bra{j \gamma}W\ket{m \beta} \right] \delta_{in} \delta_{\alpha \eta} \nonumber \\
& - \left[ \bra{j \gamma}v\ket{\beta n} - \bra{j \gamma}W\ket{n \beta} \right] \delta_{im} \delta_{\alpha \eta} \nonumber \\
& + \left[ \bra{i \gamma}v\ket{\alpha m} - \bra{i \gamma}W\ket{m \alpha} \right] \delta_{jn} \delta_{\beta \eta} \nonumber \\
& - \left[ \bra{i \gamma}v\ket{\alpha n} - \bra{i \gamma}W\ket{n \beta} \right] \delta_{jm} \delta_{\beta \eta} \nonumber \\
& - \left[ \bra{j \gamma}v\ket{\alpha m} - \bra{j \gamma}W\ket{m \gamma} \right] \delta_{in} \delta_{\beta \eta} \nonumber \\
& + \left[ \bra{j \gamma}v\ket{\alpha m} - \bra{j \gamma}W\ket{n \alpha} \right] \delta_{im} \delta_{\beta \eta} \nonumber \\
& + \left[ \bra{i \eta}v\ket{\beta m} - \bra{i \eta}W\ket{m \beta} \right] \delta_{jn} \delta_{\alpha \gamma} \nonumber \\
& - \left[ \bra{i \eta}v\ket{\beta n} - \bra{i \eta}W\ket{n \beta} \right] \delta_{jm} \delta_{\alpha \gamma} \nonumber \\
& - \left[ \bra{j \eta}v\ket{\beta m} - \bra{j \eta}W\ket{m \beta} \right] \delta_{in} \delta_{\alpha \gamma} \nonumber \\
& + \left[ \bra{j \eta}v\ket{\beta n} - \bra{j \eta}W\ket{n \beta} \right] \delta_{im} \delta_{\alpha \gamma}
\end{align}
Diagonalizing the Hamiltonian $H^{xx}$ gives excited states corresponding to double excitations. The projected Hamiltonian $H^{xx}$ has been used to calculate biexcitons in transition metal dichalcogenides\cite{Torche2021}.     \\

In molecules, the excited states are neither purely singly excited nor purely doubly excited. Some coupling between the single and double excitations leads to mixed character in the excited state. We have developed two new methods to describe such states: screened configuration interaction singles and doubles (scrCISD) and screened configuration interaction singles with perturbative doubles scrCIS(D).  

\subsection{Screened configuration interaction single and doubles (scrCISD) method}
 In the scrCISD method, single and double-excited Slater determinants are included in the basis. This combined basis is projected onto the effective Hamiltonian $H$ in Eq. (1)
 \begin{align}
     H^{scrCISD} = \begin{bmatrix}
         H^{x} & H^{x,xx} \\
         H^{xx,x} & H^{xx} \\
     \end{bmatrix}
 \end{align}
$H^{x}$ (cf. Eq. 5) is the Hamiltonian projected onto the subspace of single excitations. $H^{xx}$ (cf. Eq. 7) is the Hamiltonian projected onto the subspace of double excitations. $H^{x,xx}$ (with $H^{xx,x}$ being its Hermitian conjugate) contains the coupling between single and double excitations. 
 \begin{align}
      H^{x,xx}_{m\gamma,ij\alpha\beta} &= \bra{ij\alpha\beta} H \ket{m\gamma}  \nonumber\\
     \bra{ij\alpha\beta} H \ket{m\gamma}   &=  [ \bra{ij}W\ket{ m\alpha}-\bra{ij}W\ket{\alpha m}]\delta_{\beta \gamma} \nonumber\\
     &- [\bra{ij}W\ket{m\beta} - \bra{ij}W\ket{\beta m} ]\delta_{\alpha \gamma} \nonumber\\
 &- [\bra{\alpha \beta}W\ket{\gamma i} - \bra{\alpha \beta}W\ket{i \gamma} ]\delta_{mj} \nonumber\\
 &+ [\bra{\alpha \beta}W\ket{\gamma j} - \bra{\alpha \beta}W\ket{j \gamma}]\delta_{mi}
 \end{align}
 The Hamiltonian $H^{scrCISD}$ is diagonalized to obtain the excited states that contain both single and double excitation character. 

 \subsubsection{Including spin}
 Neglecting spin-orbit coupling, the electron and hole states can have either up spin $i\uparrow$, $\alpha \uparrow$, or down spin $i\downarrow$, $\alpha \downarrow$. Note that for the hole state, the spin corresponds to the electron that initially occupies that state. The singly excited Slater determinants denoted $\ket{i\alpha}$ have four distinct spin states.
 \begin{equation*}    
 \ket{\uparrow\uparrow}, \ket{\uparrow\downarrow}, \ket{\downarrow\uparrow}  \ket{\downarrow\downarrow}
  \end{equation*}
  The doubly excited Slater determinants denoted by $\ket{ij\alpha\beta}$ have different spin structures depending on the type of Slater determinant (SD). We divide doubly excited SDs into four groups. Group I SDs have distinct electron spatial orbitals and hole orbitals in the SD $\ket{ij\alpha \beta}$ (i.e., $i\ne j$ and $\alpha \ne \beta$). Group I SDs have 16 distinct spin states.
 \begin{align*}
     &\ket{\uparrow \uparrow \uparrow \uparrow}, \ket{\uparrow \uparrow \uparrow \downarrow}, \ket{\uparrow \uparrow \downarrow \uparrow}, \ket{\uparrow \uparrow \downarrow \downarrow},\\ 
     &\ket{\uparrow \downarrow \uparrow \uparrow}, \ket{\uparrow \downarrow \uparrow \downarrow}, \ket{\uparrow \downarrow \downarrow \uparrow}, \ket{\uparrow \downarrow \downarrow \downarrow},\\ &\ket{\downarrow \uparrow \uparrow \uparrow}, \ket{\downarrow \uparrow \uparrow \downarrow}, \ket{\downarrow \uparrow \downarrow \uparrow}, \ket{\downarrow \uparrow \downarrow \downarrow},\\ &\ket{\downarrow \downarrow \uparrow \uparrow}, \ket{\downarrow \downarrow \uparrow \downarrow},\ket{\downarrow \downarrow \downarrow \uparrow}, \ket{\downarrow \downarrow \downarrow \downarrow}
     \end{align*}
      Group II SDs have distinct electron spatial orbitals and the same hole spatial orbitals ($\ket{ij\alpha\alpha}$ with i $\ne$j) and have four distinct spin states. The Pauli exclusion principle reduces the possible spin structures to four.
      \begin{equation*}
      \ket{\uparrow \uparrow \uparrow \downarrow}, \ket{\uparrow \downarrow \uparrow \downarrow}, \ket{\downarrow \uparrow \uparrow \downarrow}, \ket{\downarrow \downarrow \uparrow \downarrow}
      \end{equation*}
      Group III SDs have distinct hole spatial orbitals and the same electron spatial orbitals ($\ket{ii\alpha\beta}$ with  $\alpha \ne \beta$) and have four distinct spin states.
      \begin{equation*}
      \ket{\uparrow \downarrow \uparrow \uparrow},  \ket{\uparrow \downarrow \uparrow \downarrow}, \ket{\uparrow \downarrow \downarrow \uparrow}, \ket{\uparrow \downarrow \downarrow \downarrow}
      \end{equation*}
      Group IV SDs have the same electron and hole spatial orbitals ($\ket{ii\alpha\alpha}$) and only have one distinct spin state $\ket{\uparrow \downarrow \uparrow \downarrow}$.\\      
The Hamiltonian commutes with all the components $S_x, S_y,$ and $S_z$ of  the total spin operator. Hence, $[H, S^{2}] = 0$ and $[H, S_{z}] = 0$, and the Hamiltonian becomes block diagonal when expressed in the eigenbasis of $S^{2}$ and $S_z$. Each block corresponds to a distinct set of eigenvalues of $S^{2}$ and $S_{z}$. This allows us to diagonalize the different reduced blocks separately rather than diagonalizing the entire Hamiltonian. We have constructed such total spin eigenstates for each type of SD separately and
denote them as $\ket{X^{m}_{n}}$, where X represents the $S^{2}$ eigenvalue (S for singlet, T for triplet and Q for quintet), $m$ corresponds to the $S_z$ eigenvalue and $n$ indeces the different states with a given ($S^2$ and $S_z$) sector (e.g., different singlets with $S^2=0$ and $S_z=0$ that belong to distinct groups).
\\
For singly excited Slater determinants (SDs), the $S^2$ eigenstates, which we refer to henceforth as  ``spin eigenstates", consist of one singlet and three triplet states (corresponding to $S_z$ = -1,0,1). For group I-doubly-excited SDs, they consist of five quintet states (corresponding to $S_z$ = -2,-1,0,1,2), nine triplet states, and two singlet states.
For group II doubly excited SDs, the spin eigenstates include one singlet and three triplet states. Similarly, group III doubly excited SDs lead to one singlet and three triplet states. Finally, group IV doubly excited SDs can only form a singlet state. The exact spin structures of the above spin eigentates are given in the Supplementary Material.
\begin{figure}[h!]
    \centering
        \includegraphics[width=\columnwidth]{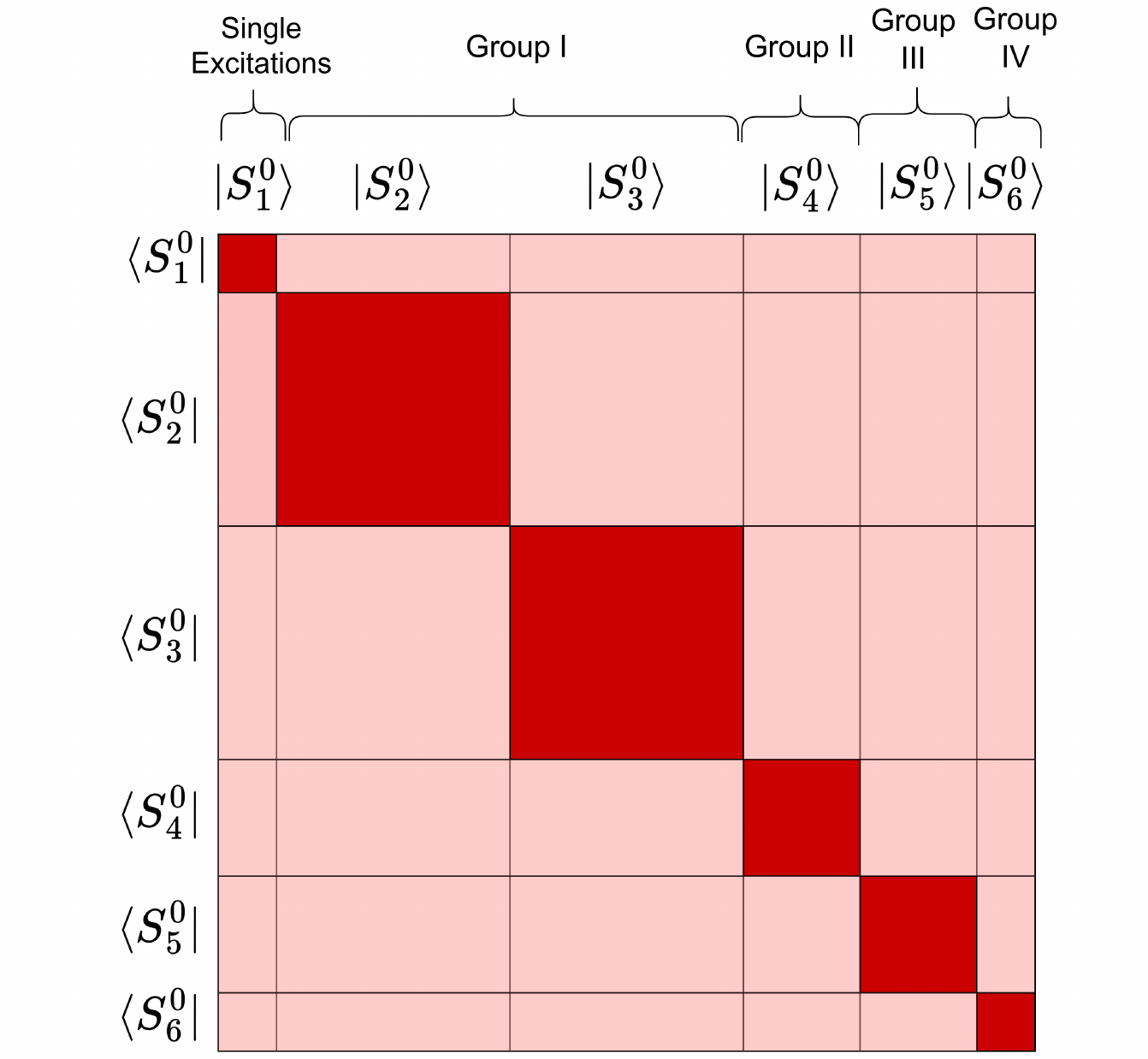}
        \caption{Singlet block of the scrCISD Hamiltonian. Each block represents singlets from a different group, and the block size qualitatively reflects the number of distinct spatial basis states in that group. Off-diagonal couplings between groups are represented in light red colour.}
\end{figure}
\begin{figure}
        \centering
        \includegraphics[scale =0.22]{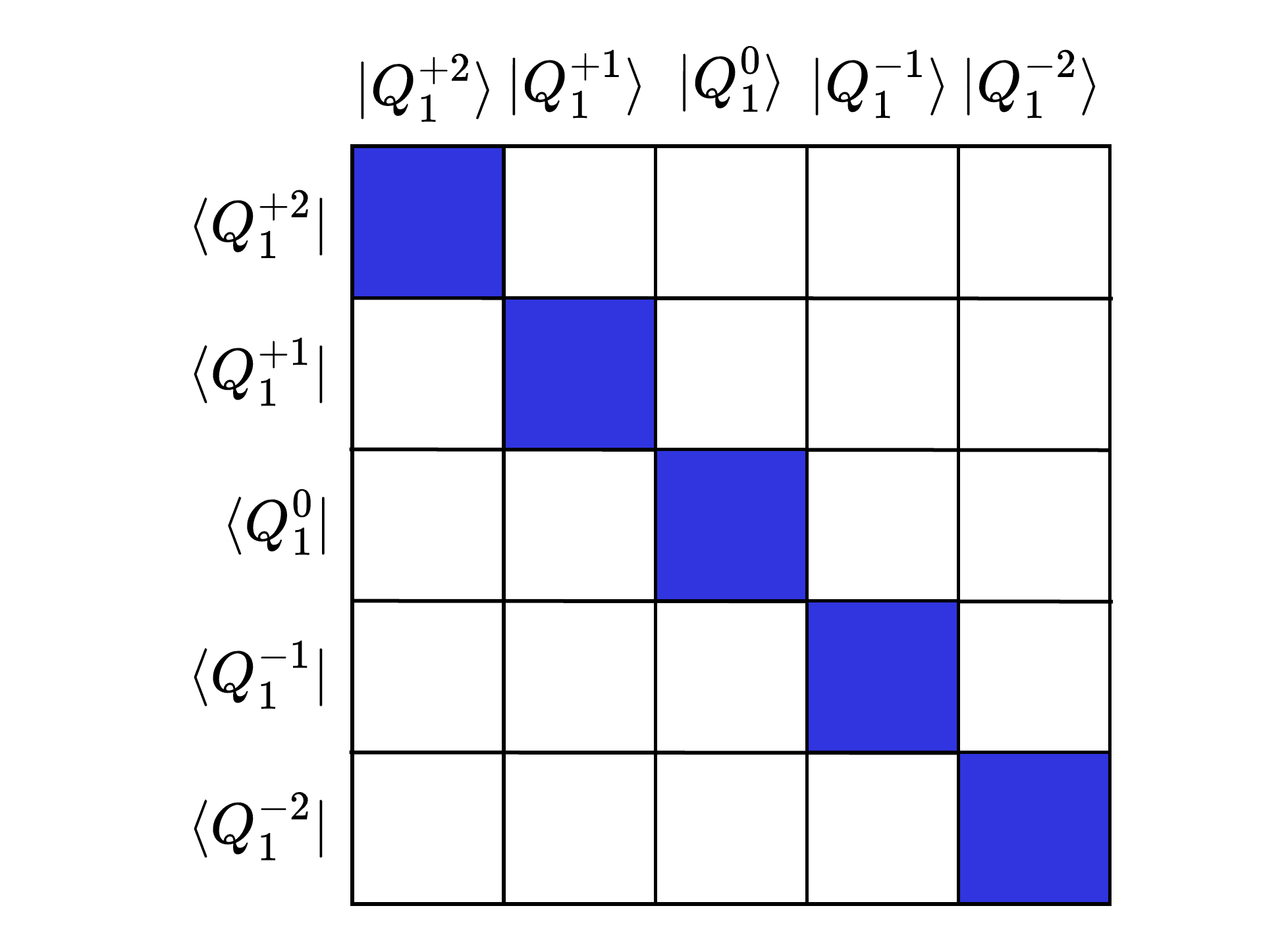}
        \caption{Quintet block of the scrCISD Hamiltonian. Each block represents a quintet state with a distinct $S_z$ eigenvalue. Because quintets with different $S_z$ values do not couple, the off-diagonal elements are zero.}
\end{figure}
Spin eigenstates can only be coupled to other spin eigenstates with the same eigenvalue for $S^2$ and $S_z$. (For example, singlets couple only with singlets and do not couple with triplets and quintets). Thus, the Hamiltonian decouples into a singlet block, three separate triplet blocks corresponding to $S_{z} = 1,0,-1$, and five separate quintet blocks corresponding to $S_{z} = 2,1,0,-1,-2$. The singlet and quintet blocks are shown schematically in Fig. 1 and Fig. 2. The singlet block is formed by one singlet from the singly excited SD, two singlets from group I doubly excited SDs, one singlet from group II doubly excited SDs, one singlet from group III doubly excited SDs, and one singlet from group IV doubly excited SDs. The diagonal quintet block only contains quintet spin eigenstates from group I.

\subsubsection{Multichannel Dyson equation for scrCISD Method (scrCISD-MCDE)}
To perform a diagrammatic analysis of the scrCISD method, we adopt the multichannel Dyson equation (MCDE) formalism\cite{PhysRevLett.131.216401,
PhysRevB.111.195133, PhysRevB.110.115140}, which couples the two-body Green's function to the four-body Green's function. Within this framework, the equations of motion for the Green's functions of the scrCISD Hamiltonian can be recast into a Dyson-type equation, which we refer to as the scrCISD-MCDE. A full derivation is provided in the Supplementary Material. The scrCISD-MCDE is given by
\begin{align}
     \Tilde{L}(\omega) &= \Tilde{L}^{0}(\omega) + \Tilde{L}^{0}(\omega)\Tilde{\Sigma}\Tilde{L}(\omega)
\end{align}
where, $\Tilde{L}$ is the interacting four-body correlation function, and $\Tilde{L^{0}}$ is the non-interacting counterpart. The non-interacting correlation function has a block-diagonal form 
\begin{align}
 \Tilde{L}^{0}(\omega) = \begin{pmatrix}
     \Tilde{L}^{0}_{2}(\omega) & 0 \\
     0 & \Tilde{L}^{0}_{4}(\omega)
 \end{pmatrix}   
\end{align} 
The $\Tilde{L}^{0}_{2}$ contains poles corresponding to one electron and one hole (two particles) channel, and the $\Tilde{L}^{0}_{4}$ block contains poles corresponding to two electrons and two holes (four particles) channel of the four-body Green's function. The self-energy $\Tilde{\Sigma}$  contains $\Tilde{\Sigma^{2p}}$ and $\Tilde{\Sigma^{4p}}$ on the diagonals that introduces correlations within the two-particle and four-particle channel respectively while the off diagonal blocks couple $\Tilde{\Sigma}^{2p,4p},\Tilde{\Sigma}^{4p,2p} $ couple the two-particle and four particle channel. Correspondingly, the interacting four-body correlation $\Tilde{L}$ contains the interacting two-body channel $\Tilde{L}^{2p}$ and interacting  four-body channel $\Tilde{L}^{4p}$ along with coupling terms, denoted as $\Tilde{L}^{2p,4p}$ and $\Tilde{L}^{4p,2p}$. Diagrammatically, the scrCISD-MCDE is shown in Fig. 3. 
\begin{widetext}
\begin{figure*}
    \centering
    \includegraphics[scale=0.14]{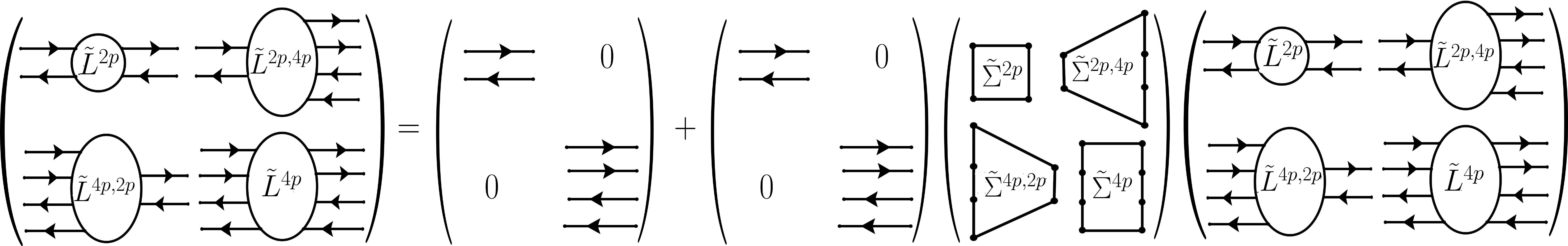}
    \caption{Diagrammatic representation of scCISD-MCDE as shown in Eq. (10)}
    \label{fig:enter-label}
\end{figure*}
\end{widetext}
By iteratively expanding the scrCISD-MCDE equation, we generate the diagrams included in the scrCISD method. The first-order diagrams obtained from Eq. (10) are shown in the Supplementary Material (SM). Figure 1 in the SM shows first-order diagrams in $\tilde{L}^{4p}$ that corresponds to the term $\Tilde{L}_2^{0}\Tilde{\Sigma}^{2p}\Tilde{L}_2^{0}$.  Figure 2 (in SM) presents the first-order diagrams in $\tilde{L}^{2p,4p}$ arising from $\Tilde{L}_2^{0}\Tilde{\Sigma}^{2p,4p}\Tilde{L}_4^{0}$. Figure 3 (in SM) shows the first-order diagrams in $\Tilde{L}^{4p}$ arising from $\Tilde{L}_4^{0}\Tilde{\Sigma}^{4p}\Tilde{L}4^{0}$. These include diagrams from both electron–electron and hole–hole interactions. In addition, there are 16 electron–hole direct interaction terms and 16 electron–hole exchange interaction terms. For clarity, only two electron–hole direct interaction diagrams and two electron–hole exchange interaction diagrams are shown. The remaining electron–hole diagrams can be drawn in an analogous manner. \\
We note that the Green's functions in Feynman diagrams have both forward and backward time orderings. However, the Green's functions appearing in the scrCISD-MCDE involve only one time ordering (as shown in the Supplementary Materials). For a Feynman diagram with $n$ time variables, all 
$n! $ time orderings are included. In contrast, each diagram in our approach corresponds to one particular time ordering. Thus, the diagrams generated within scrCISD-MCDE (as shown in Figures 1, 2, and 3 in Supplementary Material) should be interpreted as  Goldstone diagrams rather than  Feynman diagrams.  In these diagrams, time flows from left to right. Note that in Goldstone diagrams, one-body Green's function propagating along the time axis has poles associated with electrons, while one propagating opposite to the time axis has poles associated with holes.

\subsection{Screened configuration interaction singles with perturbative doubles (scrCIS(D)) method}
The screened configuration interaction singles with perturbative doubles (scrCIS(D)) method applies a perturbative energy correction to single excitation energies calculated using the scrCIS method (GW-BSE within TDA). Here, the correlation effects on the scrCIS singly excited states, arising from double and triple excitations, are incorporated via second-order perturbative corrections to the singly excited state energies. Additionally, the ground-state energy, calculated at the DFT level, is also corrected through the perturbative inclusion of doubly excited configurations. \\
 The effective Hamiltonian in Eq. (1) can be partitioned into the unperturbed Hamiltonian $H_0$ and a perturbative part $V$ that contains all the remaining interacting parts of $H$.
\begin{align}
    H = H_{0} + V
\end{align}
First, the perturbative energy correction to the DFT ground state energy from the doubly excited states ($V$ does not connect singly excited states to the ground state in accordance with Brillouin's theorem) is calculated using 2nd-order time-independent perturbation theory. The ground state amplitudes of double excitations up to first-order in perturbation is contained in the operator $T_2$    
\begin{align}
    T_2\ket{\Omega} &= \sum_{ij\alpha\beta} a^{ij}_{\alpha \beta} \ket{ij\alpha\beta} \\
    a^{ij}_{\alpha \beta} &= - \frac{\bra{ij\alpha\beta} V \ket{\Omega}}{ \Delta^{ij}_{\alpha\beta}} 
\end{align}
where $\Delta^{ij}_{\alpha\beta} = \epsilon^{qp}_{i} + \epsilon^{qp}_{j} - \epsilon^{qp}_{\alpha} -\epsilon^{qp}_{\beta} $.
The second-order energy correction $E^{2PT}_0$ is given by
\begin{align}
    E^{2PT}_{0} &= \bra{\Omega} V \ket{T_2\Omega}  
\end{align}
The singly excited state $\ket{\phi}$ (also represented in Eq. (17) as action of a single excitation operator $U_1$ acting on the ground state $\ket{\Omega}$) is calculated using scrCIS/GW-BSE within TDA.
\begin{align}
    H^{x}\ket{\phi} &= \omega\ket{\phi}  \\
    \ket{\phi} &= U_{1}\ket{\Omega} = \sum_{i\alpha} b_{\alpha}^{i} \ket{i\alpha}
\end{align}
To treat the correlations of the excited state at the same level as the ground state, the perturbative corrections from both double and triple excitations (double excitation on top of single excitation) are calculated using 2nd-order time-independent perturbation theory. The singly excited state amplitudes of double and triple excitations are contained in operators $U_2$ and $U_3$.
\begin{align}
    U_{2}\ket{\Omega} &= \sum_{ij\alpha\beta} b^{ij}_{\alpha\beta} \ket{ij\alpha\beta} \\
    U_{3}\ket{\Omega} &= \sum_{ijk\alpha\beta\gamma} b^{ijk}_{\alpha\beta\gamma} \ket{ijk\alpha\beta\gamma} \\
    b^{ij}_{\alpha\beta} &=  -\frac{\bra{ij\alpha\beta} V \ket{\phi}}{\Delta^{ij}_{\alpha\beta} - \omega} \\ 
    b^{ijk}_{\alpha\beta\gamma} &= -\frac{\bra{ijk\alpha\beta\gamma}V\ket{\phi}}{\Delta^{ijk}_{\alpha\beta\gamma} - \omega} 
\end{align}
However, the energy contribution from triple excitations is not size-consistent, meaning the energy of a system consisting of two physically separated units is not equal to the sum of their individual energies. To make this size consistent, $U_3$ is replaced by $T_2U_1$ as proposed in Ref.\cite{HEADGORDON199421}. 
Then the second-order energy correction to scrCIS state $\ket{\phi}$ is given by
\begin{align}
    E^{2PT}_{1} &=  \bra{\phi} V \ket{U_2\Omega} + \bra{\phi} V \ket{T_2U_1\Omega}   \\
    \bra{\phi} V \ket{T_2U_1\Omega} &=   \bra{\Omega} V \ket{T_2\Omega} +   \bra{\phi} V \ket{T_2U_1\Omega}_c 
\end{align}
 In  $\bra{\phi} V\ket {T_2U_1\Omega}$, the disconnected diagrams give rise to the first term in Eq. (22), which is the same as the $E^{2PT}_{0}$. The contribution from connected diagrams is 
 \begin{align}
 &\phantom{=}\bra{\phi} V \ket{T_2U_1\Omega}_c = \sum_{i\alpha} b_{\alpha}^{i} v_{\alpha}^{i} \\
   v_{\alpha}^{i} &=\frac{1}{2} \sum_{j k \beta \gamma}\bra{\beta \gamma} W \ket{j k}[b_{\alpha}^{j} a_{\beta \gamma}^{k i}+b_{\beta}^{i} a_{\alpha \gamma}^{k j}+2 b_{\beta}^{j} a_{\alpha \gamma}^{i k}]
 \end{align}
The excitation energy $\omega^{scrCIS(D)}$ is the difference between $E^{2PT}_{1}$ and $E^{2PT}_{0}$ which is given by
 \begin{align}
     \omega^{{scrCIS}(D)} &=-\frac{1}{4} \sum_{i j \alpha \beta}\frac{|u_{\alpha \beta}^{ij}|^2}{(\Delta_{\alpha \beta}^{ij}-\omega)}+\sum_{i \alpha} b_{\alpha}^{i} v_{\alpha}^{i} 
 \end{align}
 Since this is a perturbative method, it allows for the inclusion of doubly excited states without increasing the size of the Hamiltonian that is diagonalized.

\section{Computational Details}
All density functional theory (DFT), GW, and GW-Bethe-Salpeter Equation (GW-BSE) calculations were performed using the MOLGW\cite{BRUNEVAL2016149} code, which employs a Gaussian basis set. The calculations were performed on two sets of molecules.

\subsection{Thiel's Set}
A benchmark set of 28 molecules, known as Thiel's set\cite{10.1063/1.2889385}, was used to evaluate the performance of both methods. The excitation energies calculated by both methods were compared against the "Best Theoretical Estimates" (BTE)\cite{10.1063/1.2889385}, which were selected from a survey of \textit{ab initio} calculations that best match experimental results. The optimized geometries were obtained from the supplementary material of Ref\cite{10.1063/1.2889385}. Since the BTE calculations utilized the TZVP basis set\cite{10.1063/1.463096}, the same basis set was used for the scrCISD and scrCIS(D) calculations. The BHLYP exchange-correlation functional\cite{10.1063/1.464304}, which provided the best agreement for GW-BSE calculations\cite{10.1063/1.4922489}, was employed in the DFT calculations.

\subsection{Dimer Molecules}
The scrCISD method was also employed to calculate the binding energies of pentacene and tetracene dimers (Figure 10 and Figure 11), which undergo singlet fission\cite{doi:10.1021/cr1002613}. The geometry of the pentacene dimer was optimized using DFT with the CAM-B3LYP functional\cite{OKUNO201229} and a 6-31G basis set, while the geometry of the tetracene dimer was obtained from the supplementary information of the Ref\cite{doi:10.1021/jz402122m}. For the remaining calculations on both dimers, the CAM-B3LYP functional and the cc-pVTZ basis set\cite{dunning1989gaussian} were used. Electron states (i) up to 5eV above the energy of the lowest unoccupied molecular orbital and hole states ($\alpha$) up to 5eV below the energy of the highest occupied molecular orbitals were included in the scrCISD calculation.    

\section{Results and Discussion}

\subsection{Thiel's set}
 Table I in the Supplementary Materials summarizes the results of scrCISD and scrCIS(D) calculations on Thiel's set of molecules. The table also includes a column with best theoretical estimates\cite{10.1063/1.2889385} (BTE), which serves as the reference energy for calculating errors for both scrCISD and scrCIS(D) methods. The table also presents excitation energies calculated from GW-BSE calculations. The values for GW-BSE are taken from the paper by Bruneval\cite{10.1063/1.4922489} \textit{et al}. We have also reported the GW-BSE calculations done within the Tamm-Dancoff approximation (TDA). Below, we present a discussion of our key findings vis-à-vis previous work. 
\begin{center}
\begin{figure}[h]
    \centering
    \includegraphics[width=\columnwidth]{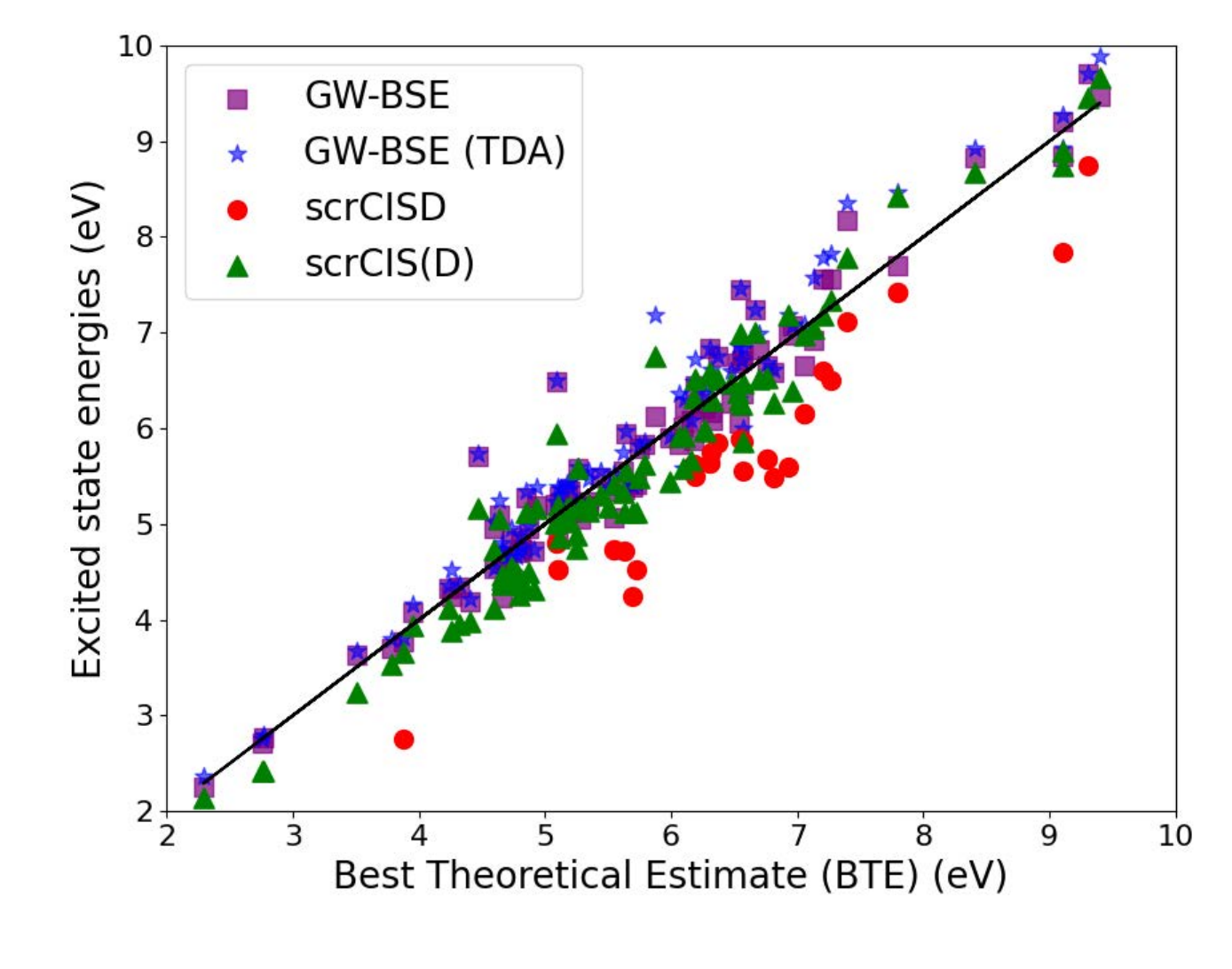}
    \caption{Correlation plot of excitation energies from GW-BSE,GW-BSE(TDA),scrCISD and scrCIS(D) method}
    \label{fig:enter-label}
\end{figure}
\end{center}
\begin{center}
\begin{figure}[h]
    \centering
    \includegraphics[width=\columnwidth]{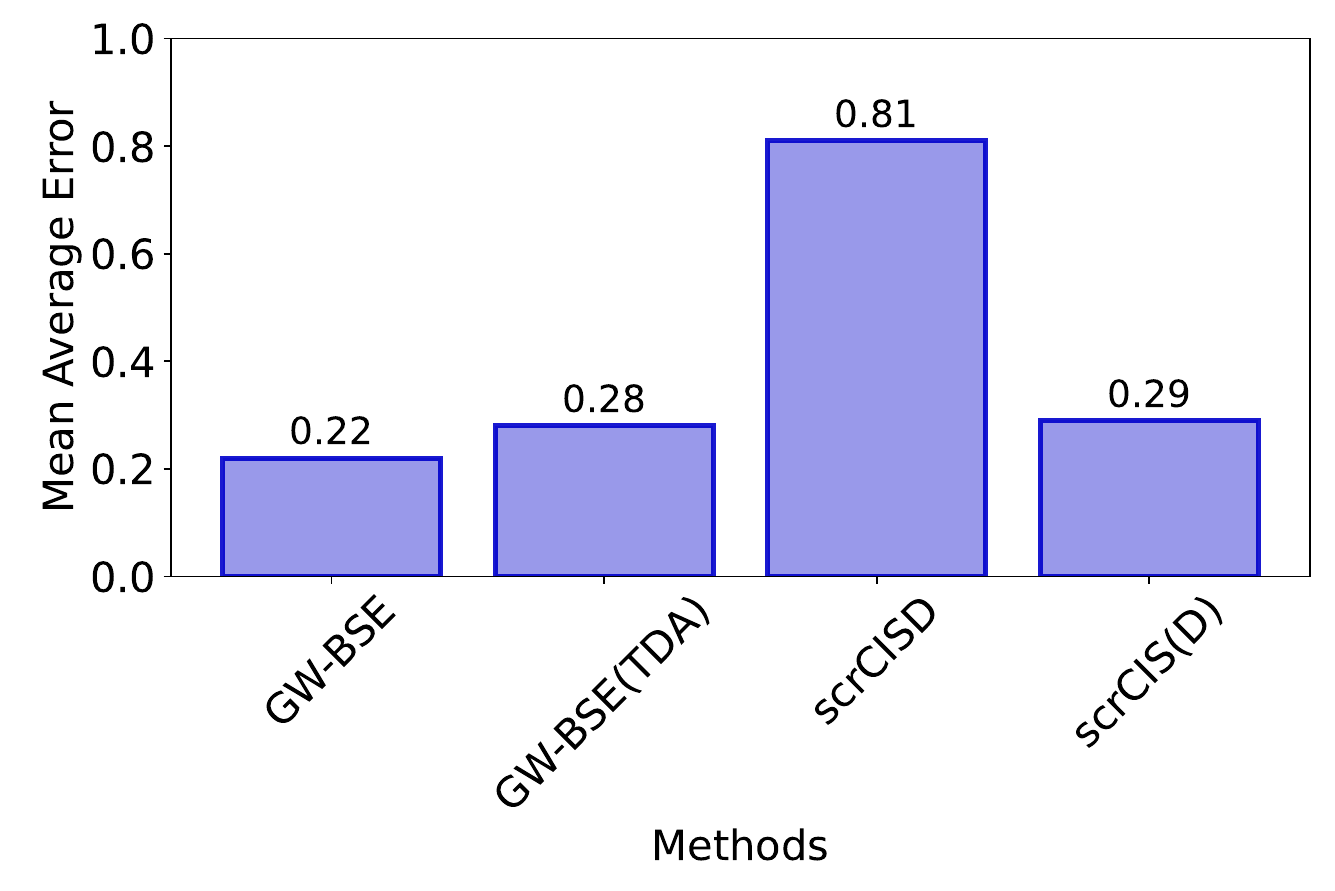}
    \caption{Bar plot of mean average errors of different methods}
    \label{fig:enter-label}
\end{figure}    
\end{center}
The correlation plot of the excitation energies from our calculations is shown in Figure 4. We note that the scrCISD method consistently underestimates the excited state energies compared to BTE. This is also reflected in a large negative value of the mean sign error (MSE) calculated for the smallest 10 molecules of Thiel's set. Most of the excited states included in Thiel's set are dominantly singly excited. This can be seen by examining the percentage of single excitation character ($\%T_1$) calculated from the CCSD\cite{10.1063/1.2889385} method, which shows most excited states have $T_1$ greater than 90\%. \\
To understand the origin of the underestimation, we analyze the diagrams included in the scrCISD Hamiltonian. The scrCISD Hamiltonian is recast into the multichannel Dyson equation called scrCISD-MCDE as given by Eq. (10). Upon iterative expansion of the scrCISD-MCDE, the interacting two-body correlation function $\Tilde{L}^{2p}$ is given by
\begin{align}
      \Tilde{L}^{2p}&= \Tilde{L}_2^{0}+ \Tilde{L}_2^{0}\Tilde{\Sigma}^{2p}\Tilde{L}_2^{0}+\Tilde{L}_2^{0}\Tilde{\Sigma}^{2p}\Tilde{L}_2^{0}\Tilde{\Sigma}^{2p}\Tilde{L}_2^{0}\\ \nonumber
      &\phantom{=}+ \Tilde{L}_2^{0}\Tilde{\Sigma}^{2p,4p}\Tilde{L}_4^{0}\Tilde{\Sigma}^{4p,2p}\Tilde{L}_2^{0} + \dots
\end{align}

\begin{figure}[h]
  \centering
  \subfigure[]{\includegraphics[width=0.49\linewidth]{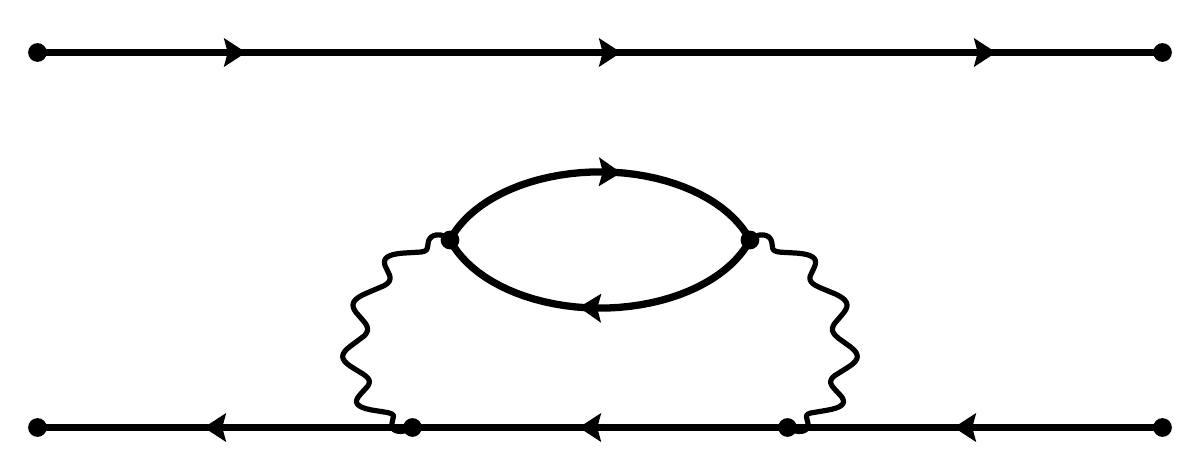}}
  \hfill
  \subfigure[]
  {\includegraphics[width=0.49\linewidth]{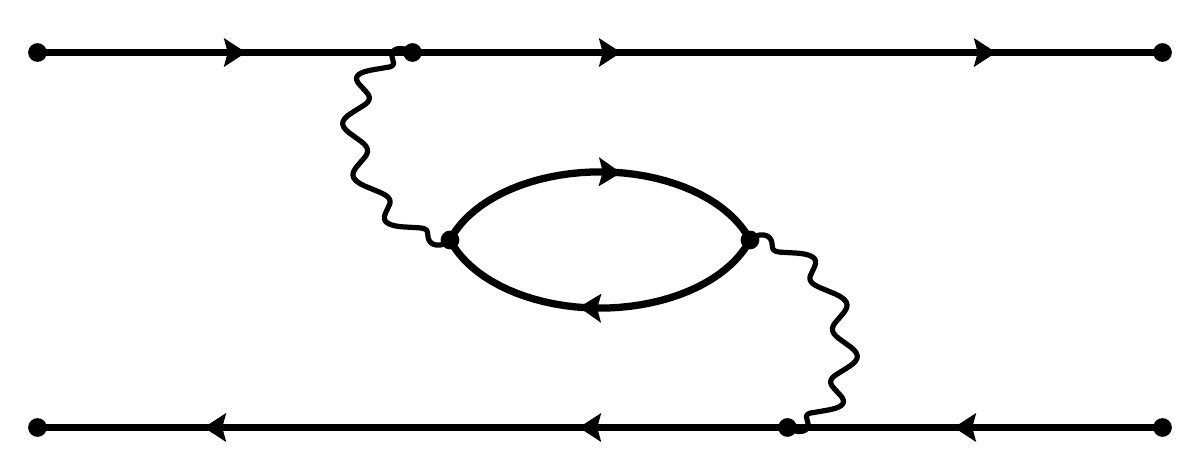}}
  \caption{Second-order diagrams that are generated from the coupling of the two-body correlation function $\Tilde{L}^{0}_{2}$ with the four-body correlation function $\Tilde{L}^{0}_{4}$ in scrCISD-MCDE}. 
  \label{fig:combined}
\end{figure}
\begin{figure}[h]
    \centering
    \includegraphics[width=\linewidth]{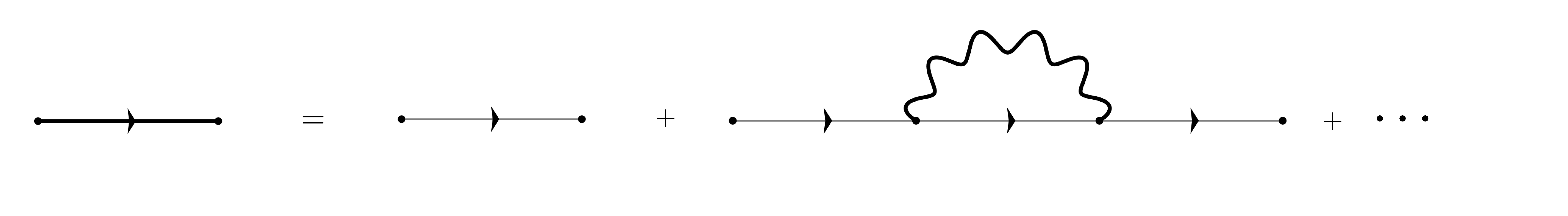}
    \caption{The one-body Green's function calculated within the GW approximation (thick black line) is expanded in terms of the Hartree-Fock Green's function (thin grey line) by using the Dyson equation for one-body Green's function.}
    \label{fig:placeholder}
    \includegraphics[width=0.8\linewidth]{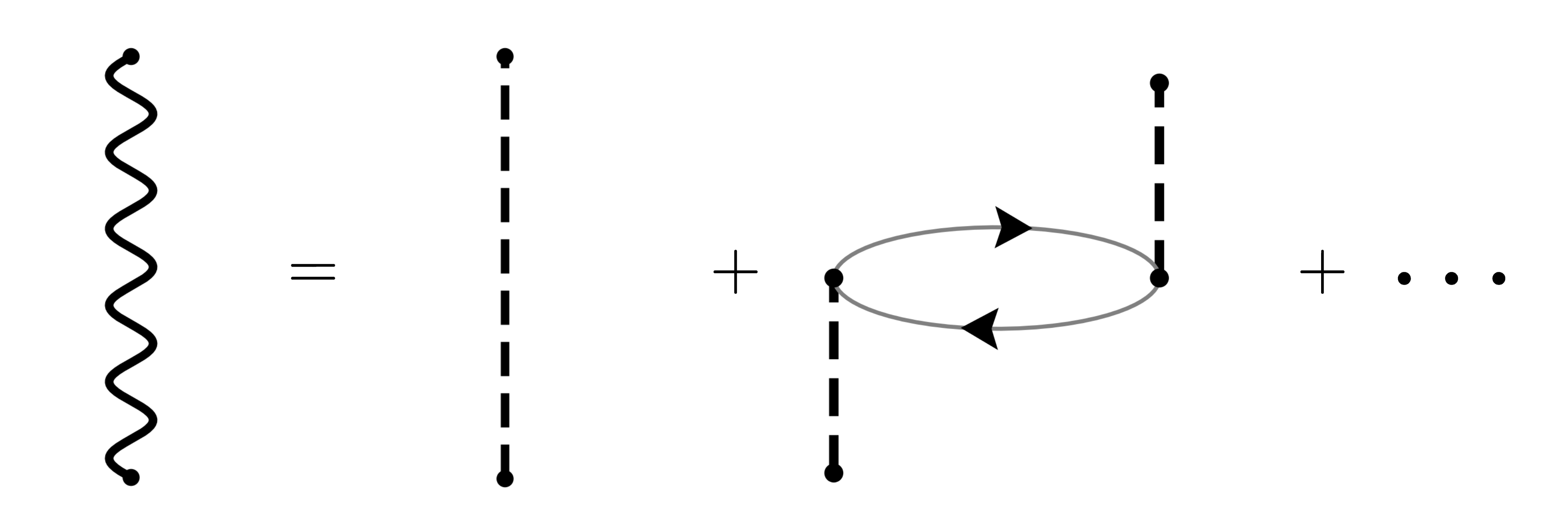}
    \caption{The screened Coulomb interaction (solid wiggly line) calculated within random phase approximation is expressed as a Dyson series expansion in terms of bare Coulomb interaction (dashed line).}
    \label{fig:placeholder}
\end{figure}
Figures 6(a) and 6(b) show second-order diagrams included in the fourth term of Eq. (27). Figure 6(a) arises from the coupling between the two- and four-body correlation functions, generating one-body self-energy diagrams analogous to those in the GW approximation. In scrCISD-MCDE, the one-body Green's functions are already constructed using the GW self-energy, whereas in MCDE, the Hartree-Fock Green's function is used. Moreover, the interaction lines in the figure correspond to screened Coulomb interactions, in contrast to the bare Coulomb interactions used in MCDE. By replacing the GW Green's function with the Hartree-Fock Green's function using the diagrammatic expansion shown in Figure 7, and substituting the screened interaction lines with bare Coulomb interactions as shown in Figure 8, one can show that the resulting diagram is already included in the first term of Eq. (27), thereby leading to double counting.

Figure 6(b) shows a diagram where the coupling of two- and four-body correlations leads to a bubble insertion between interaction lines, effectively screening the interaction (within the random phase approximation). However, since the direct electron-hole interaction is already screened in the two-body self-energy $\Tilde{\Sigma}^{2p}$, such a diagram is already included in the second term in Eq. (27). Therefore, including it again leads to double-counting.\\
\begin{figure}[h]
    \centering
    \subfigure[]{
    \includegraphics[width=0.60\linewidth]{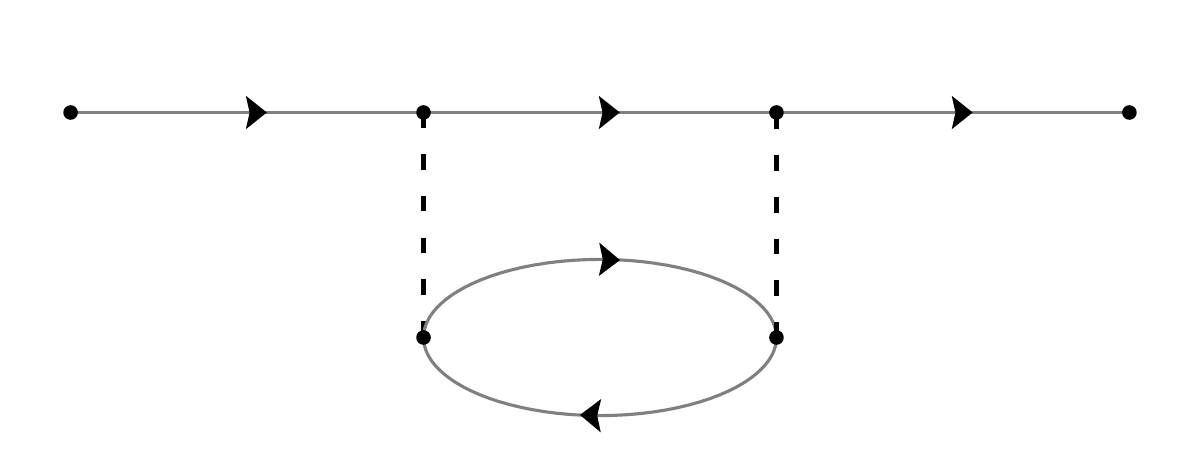}}
    \hfill
    \subfigure[]{
    \includegraphics[width=0.35\linewidth]{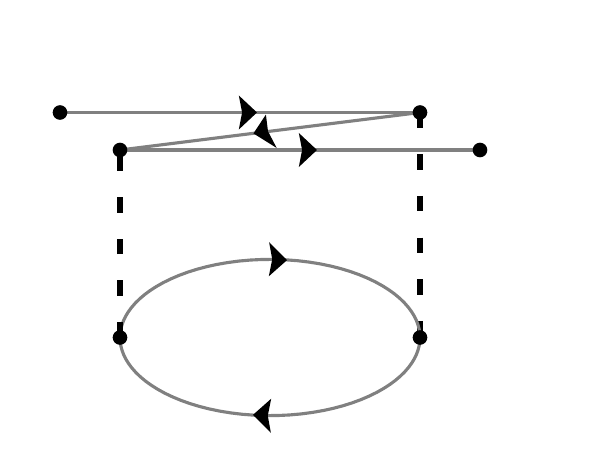}}
    \caption{Second-order Goldstone diagrams included in the GW approximation.}
    \label{fig:placeholder}
\end{figure}
Next, to eliminate the double counting of diagrams, we replaced the GW  quasiparticle energies in $\Tilde{L}^{0}_{2}$ with the Hartree-Fock energies. We replaced the screened Coulomb interaction in $\Tilde{\Sigma}^{2p},\Tilde{\Sigma}^{2p,4p}$ and $\Tilde{\Sigma}^{4p,2p}$ with bare Coulomb interaction. However, this only further worsened the results (values not reported). This is because, even though the double-counting issue is eliminated, the single particle self-energy diagrams generated through the coupling of two and four-particle correlation functions only include one particular set of Goldstone diagrams rather than all possible Goldstone diagrams included in the Feynman diagrams corresponding to the GW approximation. For example, Figures 9(a) and 9(b) are the two Goldstone diagrams corresponding to the second-order Feynman diagram for the GW self-energy. Because of the way the MCDE is constructed with time restrictions, it can only generate one of the Goldstone diagrams (Figure 9(a)) without including the other one (Figure 9(b)). This removal of essential diagrams included in the GW calculations in scrCISD-MCDE leads to larger errors in the modified scrCISD calculation. Bintrim\cite{Bintrim2022-mv} \textit{et al.} have commented on obtaining underestimated energies when using a Hamiltonian similar to the modified scrCISD Hamiltonian (Eq. 14 in Ref\cite{Bintrim2022-mv}).  In their analysis, the Goldstone diagram shown in Figure 9(a) represents an electron state renormalized via a two-electron–one-hole configuration, while Figure 9(b) corresponds to renormalization via a two-hole–one-electron process. They emphasized that retaining only one of these two diagrams, as in the modified scrCISD-MCDE formulation due to time-ordering constraints, can severely affect the GW quasiparticle energies, which in turn affect the excitation energies.  

Also, the inclusion of de-excitation energies, as done in MCDE, results in the inclusion of ground-state correlation effects. As shown by Li \textit{et al.} \cite{10.21468/SciPostPhys.8.2.020}, the two-body correlation function approximated at the Tamm-Dancoff approximation (TDA) (contains poles only corresponding to single excitations) does not introduce any correlation to the DFT ground state. Whereas when the two-body correlation function is treated within the random phase approximation (including poles corresponding to both excitation and de-excitation), this leads to additional energy terms that add to the ground state correlation. Thus, the scrCISD method also neglects the effect of ground-state correlation on the excited-state energies. \\ 
The combined effect of diagrammatic double-counting and the absence of ground-state correlation contributes to the systematic underestimation of excitation energies in scrCISD. At the scrCIS level (GW-BSE within TDA), the absence of ground-state correlation does not introduce significant error, because both the ground and excited states are treated at the same (zeroth-order) level. However, in scrCISD, the excited states benefit from correlation effects via coupling to double excitations, resulting in the lowering of their energies, while the ground state remains uncorrelated. Consequently, the excitation energies, defined as the difference between the two, are underestimated.

An alternate way to incorporate correlation effects in the ground state is to explicitly
allow it to mix with higher excited states. Although the ground state does not couple with single excitations (due to Brillouin's theorem), it does couple strongly with double excitations. Achieving a balanced description of both the ground state and the singly excited states would therefore require including triple excitations to calculate the correlation effects on the singly excited states. This would require one to use computationally expensive screened configuration interaction single doubles and triples method (scrCISDT).

The scrCIS(D) method, instead, allows for perturbative inclusion of higher excitations up to triple excitation for the excited states and up to double excitations for the ground state. This allows for a balanced description of the excited and ground states, unlike scrCISD. The scrCIS(D) performs much better (mean average error 0.29) compared to the scrCISD (mean average error 0.81) method. The scrCIS(D) MAE is comparable to the mean average error of the GW-BSE method (Fig. 4). For some states with significant double excitation character (as shown in Table I), the scrCIS(D) method performs better than GW-BSE. Thus, scrCIS(D) better describes states with double excitation character compared to GW-BSE while not compromising the Thiel's set's overall accuracy. 
  \begin{table}[h]
      \centering
       \caption{Results of states with significant double excitation character in Thiel's set}
      \label{tab:my_label}
      \begin{ruledtabular}
      \begin{tabular}{cccccc}
      \\
      Molecule  & \makecell{Sym-
      \\metry} & BTE & GW-BSE & \makecell{GW-BSE\\(TDA)} & \makecell{scr-\\
      CIS(D)}  \\
      \hline
      \\
     E-Butadiene  & $A_{g}$ & 6.55 & 7.45 &7.47 & 6.99 \\
     all-E-Hexatriene & $A_{g}$ & 5.09 & 6.49 & 6.50 &  5.94     \\
     all-E-Octatetraene &  $A_g$ & 4.47 & 5.71   & 	5.73 & 	5.16\\
     Cyclopentadiene & $A_{1}$ &  6.31 & 6.83 & 6.83 &	6.58  \\  
      Benzene           &  $E_{1g}$& 8.41& 8.83    & 8.92 &  8.68 \\
      \end{tabular}
     \end{ruledtabular}
  \end{table}

\subsection{Dimer molecules}
\begin{figure}[h!]
    \centering

        \centering
        \includegraphics[width=\columnwidth]{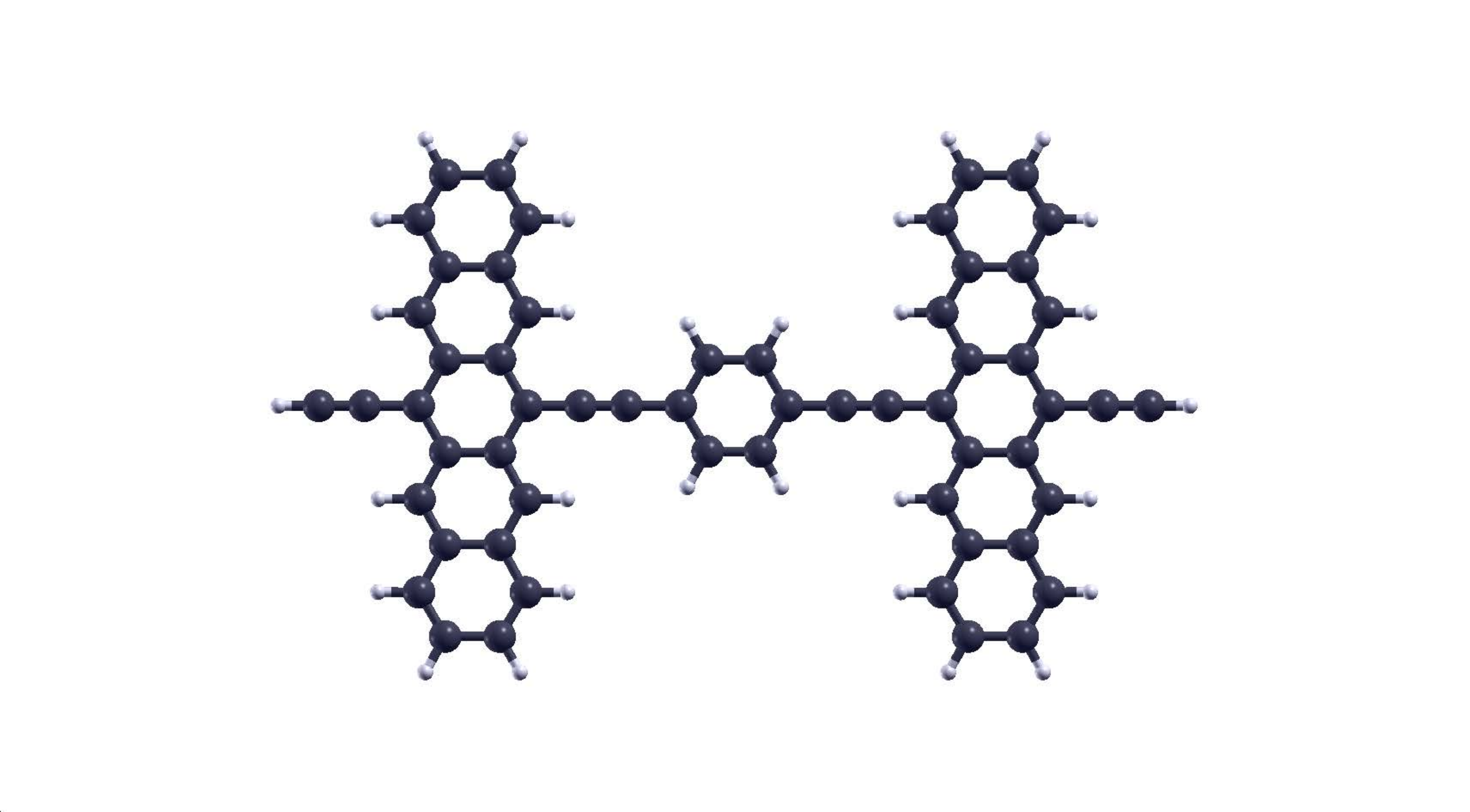}
        \caption{Pentacene dimer connected by a phenyl bridge}
\end{figure}
\begin{figure}
        \centering
        \includegraphics[width = \columnwidth]{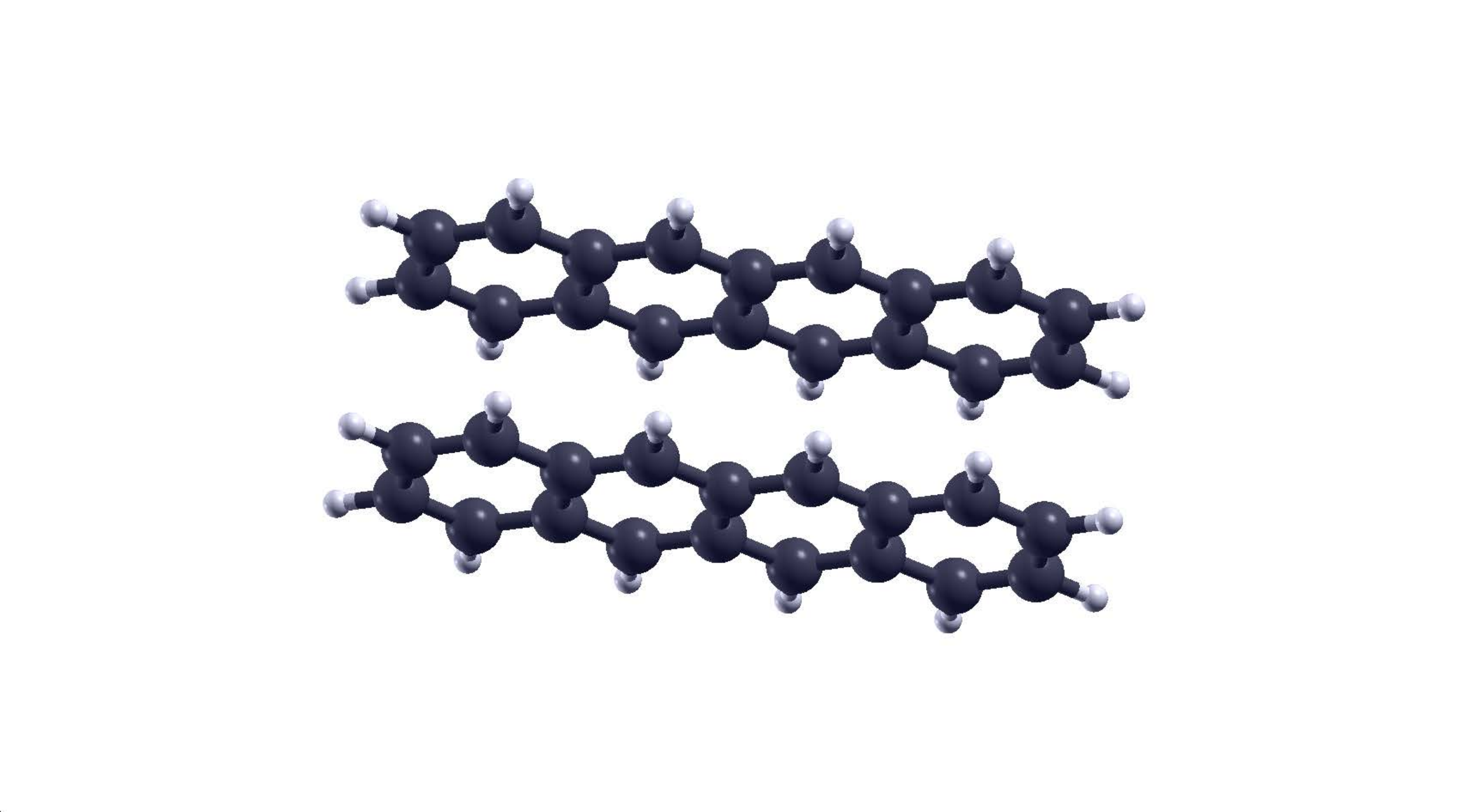}
        \caption{Tetracene dimer separated by distance of 3.7\AA}

\end{figure}
\begin{table}[h!]
        \centering
          \caption{Pentacene dimer calculation}
        \begin{ruledtabular}
        \begin{tabular}{c c c}
        \\
        
           state  & scrCISD & MRSDCI\cite{doi:10.1021/acs.jpcc.9b09831}\\
           \hline
           \\

           $(\mathrm{TT})^1$  &  1.67    &  1.71         \\
           $(\mathrm{TT})^5$ &   1.72     & 1.76         \\
           B.E.   & 0.05  & 0.05 \\
        \end{tabular}
        \end{ruledtabular}
      
\end{table}
      \begin{table}[h!]
      \caption{Tetracene dimer calculation}
        \centering
        \begin{ruledtabular}
        \begin{tabular}{ccc}
        \\
           state  & scrCISD & RAS-2SF\cite{doi:10.1021/jz402122m}  \\
           \hline
           \\
           $(\mathrm{TT})^1$  & 2.43      &   2.74   \\
           $(\mathrm{TT})^5$  &  2.66     &   2.96   \\
           B.E      &  0.23      &  0.22 \\
        \end{tabular}
        \end{ruledtabular}        
    \end{table}
We have also used the scrCISD method to describe dominantly doubly excited states. Such states are found in molecules undergoing singlet fission\cite{doi:10.1021/cr1002613}, a process in which a high-energy singlet state converts into two low-energy triplet states. This process involves an intermediate doubly excited state formed by two interacting triplet excitations residing in neighboring molecules. Two triplet excitations can interact to form either a singlet $\mathrm{(TT)^1}$, a triplet $\mathrm{(TT)}^3$ or a quintet $\mathrm{(TT)^5}$. The biexcitonic binding energy between the triplets is calculated as the difference between the singlet $\mathrm{(TT)^1}$ and the quintet state $\mathrm{(TT)^5}$. The scrCISD method is used to calculate the biexcitonic binding energies of such molecules. \\
For dominantly doubly excited states, the relevant diagrams reside in the four-body correlation function, $\tilde{L}^{4p}$. By iteratively expanding the scrCISD-MCDE equation (Eq. (10)), $\tilde{L}^{4p}$ can be expressed as 
\begin{align}
\tilde{L}^{4p} &= \tilde{L}_4^{0} + \tilde{L}_4^{0} \tilde{\Sigma}^{4p} \tilde{L}_4^{0} + \tilde{L}_4^{0} \tilde{\Sigma}^{4p} \tilde{L}_4^{0} \tilde{\Sigma}^{4p} \tilde{L}_4^{0} \nonumber \\
&\quad + \tilde{L}_4^{0} \tilde{\Sigma}^{4p,2p} \tilde{L}_{2}^{0} \tilde{\Sigma}^{2p,4p} \tilde{L}_4^{0} + \dots
\end{align}
In dominantly singly excited states, the second-order terms in $\tilde{L}^{2p}$ that arise from the coupling between the two-body and four-body correlation functions (fourth term in Eq. (27)) can lead to double counting. This occurs in two ways. The first is by adding a one-body self-energy to an already renormalized Green’s function (Figure 6(a)). The second is by inserting a polarization bubble into an already screened interaction (Figure 6(b)). However, the second-order terms in Eq. (28) do not renormalize the one-body Green’s function, nor do they further screen the interaction lines. Thus, no double counting occurs in this case.
Also, unlike the case of Thiel's set of molecules, the scrCISD method doesn't suffer from the unbalanced treatment of ground and excited states (arising due to exclusion of ground state correlation)  when used to treat dominantly doubly excited states. Since both the doubly excited and the ground state are treated at zeroth order without any correlation effects from higher excited states, this method performs well when applied to a dominantly doubly excited state.\\
To illustrate this, we have performed the scrCISD calculation on a pentacene dimer (as shown in Figure 10). The multiple reference singles and doubles CI (MRSDCI)\cite{doi:10.1021/acs.jpcc.9b09831} calculation on an empirical  Pariser–Parr–Pople Hamiltonian\cite{10.1063/1.1699030} was used to calculate the biexcitonic binding energies of the pentacene dimers. As shown in Table II, the scrCISD results agree well with the MRSDCI results. For the tetracene dimer (Figure 11), Feng\cite{doi:10.1021/jz402122m} \textit{et al.} calculated the biexcitonic binding energies of using \textit{ab initio} restricted active space double spin-flip\cite{C2CP43293E} (RAS-2SF). The biexcitonic binding energy calculated from scrCISD agrees with the results obtained from the RAS-2SF method (Table III).   

\section{Conclusions}

In conclusion, we have developed two methods—scrCISD and scrCIS(D)—to describe doubly excited states in molecular systems. Our results show that the scrCISD method systematically underestimates excitation energies for Thiel's set of molecules, which are predominantly characterized by singly excited states. This underestimation stems from two main factors: (i) double counting of diagrams and (ii) the neglect of ground-state correlation energy.

Notably, attempts to remove double counting by modifying the scrCISD Hamiltonian lead to even worse performance. A diagrammatic analysis reveals that scrCISD fails to include several Goldstone diagrams present in the Feynman diagram of the GW self-energy. Additionally, scrCISD does not treat ground and excited states at a consistent level of correlation, further contributing to the inaccuracy in excitation energies.

In contrast, the scrCIS(D) method addresses these limitations and yields accurate excitation energies for dominantly singly excited states in Thiel’s benchmark set. However, for states with significant double excitation character—as observed in singlet fission chromophores—the scrCISD method performs well, without suffering from the imbalanced treatment of correlation effects seen in singly excited states.

\section{ACKNOWLEDGEMENTS}

We thank the Supercomputer Education and Research Centre (SERC) at the Indian Institute of Science (IISc) for providing the computational
facilities. M.J. and H.R.K. gratefully acknowledge the Nano mission of the Department of Science and Technology, India, and the Indian National Science Academy, India, for financial support under Grants No. DST/NM/TUE/QM-10/2019 and No. INSA/SP/SS/2023/, respectively.
\nocite{*}
\bibliography{ref}

\end{document}